\documentclass[preprint,superscriptaddress,aps]{revtex4}
\usepackage{amsfonts}
\usepackage{amssymb}
\usepackage{amsmath}
\usepackage{graphics}
\usepackage{graphicx}
\usepackage{epsf}
\usepackage{color}
\usepackage[T1]{fontenc}
\usepackage{hyperref}
\newcommand{\bea}{\begin{eqnarray}}
\newcommand{\eea}{\end{eqnarray}}

\begin{document}
\title{Superfield Effective Potential for the Supersymmetric Topologically Massive Gauge theory in Four Dimensions}

\author{F. S. Gama}
\email{fgama@fisica.ufpb.br}
\affiliation{Departamento de F\'{\i}sica, Universidade Federal da Para\'{\i}ba\\
 Caixa Postal 5008, 58051-970, Jo\~ao Pessoa, Para\'{\i}ba, Brazil}

\author{M. Gomes}
\email{mgomes@fma.if.usp.br}
\affiliation{Departamento de F\'{\i}sica Matem\'atica, Universidade de S\~ao Paulo,\\
Caixa Postal 66318, 05314-970, S\~ao Paulo, SP, Brazil}

\author{J. R. Nascimento}
\email{jroberto@fisica.ufpb.br}
\affiliation{Departamento de F\'{\i}sica, Universidade Federal da Para\'{\i}ba\\
 Caixa Postal 5008, 58051-970, Jo\~ao Pessoa, Para\'{\i}ba, Brazil}

\author{A. Yu. Petrov}
\email{petrov@fisica.ufpb.br}
\affiliation{Departamento de F\'{\i}sica, Universidade Federal da Para\'{\i}ba\\
 Caixa Postal 5008, 58051-970, Jo\~ao Pessoa, Para\'{\i}ba, Brazil}

\author{A. J. da Silva}
\email{ajsilva@fma.if.usp.br}
\affiliation{Departamento de F\'{\i}sica Matem\'atica, Universidade de S\~ao Paulo,\\
Caixa Postal 66318, 05314-970, S\~ao Paulo, SP, Brazil}

\begin{abstract}
We explicitly calculate the one-loop k\"{a}hlerian effective potential for the supersymmetric topologically massive gauge theory in four dimensions which involves two gauge superfields, the usual scalar one and the spinor one originally introduced by Siegel, coupled to a chiral scalar matter.
\end{abstract}

\maketitle

\section{Introduction}

It is well known (see f.e. \cite{SGRS}) that in a four-dimensional space-time, three types of constrained superfields exist, consisting of irreducible representations of the supersymmetry algebra, that is, chiral, antichiral and linear superfields. The scalar chiral and antichiral superfields are well studied being the basic ingredients of the Wess-Zumino model and many other field theories \cite{BuKu,ourcourse}. Another well-studied important example is a real gauge superfield which is a reducible one since it is unconstrained, represents itself as a natural superfield extension of the usual gauge field, being thus a basic ingredient for supergauge theories such as super-QED and super-Yang-Mills theory (for different aspects of supergauge theories, see \cite{SGRS,BuKu,ourcourse} and many other textbooks). However, these models do not exhaust the set of physically interesting theories. 

In this paper, we consider a model for a spinor chiral superfield coupled to an usual chiral matter. Originally, the spinor chiral superfield was introduced in \cite{Siegel} where it was shown to correspond to the so-called tensor multiplet and to allow for introducing, first, a new supergauge model, and second, a topological mass term in the case of coupling of the spinor gauge superfield to the usual real gauge scalar superfield. One more interesting feature of this model is that the gauge invariant strength, corresponding to spinor chiral and antichiral superfields, is just a linear superfield, differently from the chiral one, occurring for the real scalar superfield \cite{SGRS}. While in \cite{Siegel} only the free theory has been considered, we study here its coupling to a chiral matter. Classical aspects of this model were discussed in \cite{Fe}. An alternative coupling for the linear superfield and tensor multiplet has been discussed in \cite{JL}, where some of its string-related aspects were considered (for applications of this multiplet see also the references therein). Using the previously developed superfield effective potential methodology \cite{WZ,WZ1,GC,SYM}, we calculate the one-loop superfield effective potential for this theory. We emphasize that, up to now, there were no examples of quantum calculations involving the chiral spinor superfields.

The structure of the paper looks like follows. In the section II, we discuss the classical action of the chiral spinor gauge superfield, coupled to the usual scalar gauge superfield and a chiral matter. In the section III we calculate the one-loop effective potential in this theory, and  the section IV contains the summary of our results.

\section{Supersymmetric topologically massive gauge theory}

Let us start our study with the supersymmetric topologically massive gauge theory which will be used to find the one-loop K\"{a}hlerian effective potential (KEP). In the pure gauge sector, we have \cite{Siegel}
\bea
\label{puregauge}
S_{G}=\frac{1}{2}\int d^6zW^\alpha W_\alpha-\frac{1}{2}\int d^8zG^2-m\int d^8zVG \ ,
\eea
where $m$ is a constant with mass dimension equal to 1, and
\bea
\label{WG}
W_\alpha=i\bar D^2D_\alpha V \ , \ G=-\frac{1}{2}(D^\alpha\psi_\alpha+\bar D^{\dot{\alpha}}\bar\psi_{\dot{\alpha}}) \ ,
\eea
where $\psi_{\alpha}$, $\bar{\psi}_{\dot{\alpha}}$ are chiral and antichiral spinor superfield corresponding to the tensor multiplet \cite{SGRS}, and $V$ is an usual real gauge superfield. Actually, $G$ is a linear superfield satisfying the relation $D^2G=\bar{D}^2G=0$.
The superfield strengths $W_\alpha$, $G$, and the action (\ref{puregauge}) are invariant under the Abelian gauge transformations:
\bea
\label{gaugetrans}
\delta V=i(\bar\Lambda-\Lambda) \ , \ \delta\psi_\alpha=i\bar D^2D_\alpha L \ , \ \delta\bar\psi_{\dot{\alpha}}=-iD^2\bar D_{\dot{\alpha}}L \ ,
\eea
where $\Lambda$ is a chiral superfield, and $\bar\Lambda$ is an antichiral one, and $L=\bar L$ is a real scalar one \cite{SGRS}.

Let us show that the theory (\ref{puregauge}) describes a massive gauge theory. For this, let us extract the equations of motion by varying the action (\ref{puregauge}) with respect to the superfields $V$ and $\psi_\alpha$. Then, we get
\bea
\label{eqW}
\frac{\delta S_{G}}{\delta V}&=&iD_\alpha W^\alpha-mG=0 \ ,\\
\label{eqG}
\frac{\delta S_{G}}{\delta \psi_\alpha}&=&\bar D^2D^\alpha G-imW^\alpha=0 \ .
\eea
On the one hand, if we multiply eq. (\ref{eqW}) by $\bar D^2D^\beta$ and use $\bar D^2D^\beta D_\alpha W^\alpha=\Box W^\beta$, we get
\bea
\label{KGW}
(\Box-m^2)W^\alpha=0 \ .
\eea
On the other hand, if we multiply eq. (\ref{eqG}) by $D_\alpha$ and use $D_\alpha\bar D^2 D^\alpha G=\Box G$, we get
\bea
\label{KGG}
(\Box-m^2)G=0 \ .
\eea
Therefore, from eqs. (\ref{KGW}) and (\ref{KGG}), we can conclude that the superfield strengths $W_\alpha$ and $G$ satisfy massive Klein-Gordon equations.

In order to perform quantum calculations, we must add to (\ref{puregauge}) a gauge-fixing term. In particular, we will consider the following one \cite{Siegel}:
\bea
\label{gaugefix}
S_{GF}=-\frac{1}{2\alpha}\int d^8zV\{D^2,\bar D^2\}V-\frac{1}{8\beta}\int d^8z(D^\alpha\psi_\alpha-\bar D^{\dot{\alpha}}\bar\psi_{\dot{\alpha}})^2 \ ,
\eea
where $\alpha$ and $\beta$ are the gauge-fixing parameters. The ghosts are completely factorized since the theory is Abelian.

Now, let us introduce interaction between the (anti-)chiral scalar superfield and the gauge superfields \cite{BuKu}. Under the usual gauge transformation, the chiral and antichiral matter superfields transform as \cite{SGRS}
\bea
\label{mattertrans}
\Phi^{\prime}=e^{2ig\Lambda}\Phi \ , \  \bar\Phi^{\prime}=\bar\Phi e^{-2ig\bar\Lambda} \ .
\eea
The interaction term that we will consider in this paper, which is invariant under the combined transformations (\ref{gaugetrans}) and (\ref{mattertrans}), is given by \cite{Chris}
\bea
\label{matteract}
S_M=\int d^8z\bar\Phi e^{2gV}\Phi e^{4hG} \ .
\eea
The coupling constants $g$ and $h$ have mass dimensions zero and $-1$, respectively. The model $S_G+S_M$ [see eqs. (\ref{puregauge}) and (\ref{matteract})] was considered in \cite{Fe} in the study of the formation of cosmic strings.

It follows from this expression that the tree-level KEP is
\bea
K^{(0)}=\Phi\bar{\Phi}.
\eea

Finally, the supersymmetric topologically massive gauge theory that we will study in this work follows from (\ref{puregauge}), (\ref{gaugefix}), and (\ref{matteract}):
\bea
\label{totaltheory}
S&=&-\frac{1}{2}\int d^8z V(-D^\alpha \bar D^2D_\alpha+\frac{1}{\alpha}\{D^2,\bar D^2\})V-\frac{1}{8}\int d^8z \Big\{\big(1+\frac{1}{\beta}\big)[\psi_\alpha D^\alpha D^\beta\psi_\beta\nonumber\\
&&+\bar\psi_{\dot{\alpha}}\bar D^{\dot{\alpha}}\bar D^{\dot{\beta}}\bar\psi_{\dot{\beta}}]+2\big(1-\frac{1}{\beta}\big)\psi_\alpha D^\alpha\bar D^{\dot{\beta}}\bar\psi_{\dot{\beta}}\Big\}+\frac{m}{2}\int d^8zV(D^\alpha\psi_\alpha+\bar D^{\dot{\alpha}}\bar\psi_{\dot{\alpha}})\nonumber\\
&&+\int d^8z\bar\Phi e^{2gV}\Phi e^{-2h(D^\alpha\psi_\alpha+\bar D^{\dot{\alpha}}\bar\psi_{\dot{\alpha}})} \ ,
\eea
where we explicitly wrote the gauge superfields.

The standard method of calculating the effective action is based on the methodology of the loop expansion \cite{BO}. To do this, we make a shift $\Phi\rightarrow\Phi+\phi$ in the superfield $\Phi$ (together with the analogous shift for $\bar\Phi$), where now $\Phi$ is a background (super)field and $\phi$ is a quantum one. We assume that the gauge superfields $V$, $\psi_\alpha$, and $\bar\psi_{\dot{\alpha}}$ are quantum. In order to calculate the effective action at the one-loop level, we have to keep only the quadratic terms in the quantum superfieds. By using this prescription, we get from (\ref{totaltheory})
\bea
\label{backaction1}
&&S_2[\bar\Phi,\Phi;\bar\phi,\phi,\psi_\alpha,\bar\psi_{\dot{\alpha}},V]=S_q+S_{int} \ ,\\
&&S_q=\frac{1}{2}\int d^8z\big[ -V\Box(\Pi_{1/2}+\frac{1}{\alpha}\Pi_0)V-\frac{1}{4}\big[\big(1+\frac{1}{\beta}\big)(\psi_\alpha D^\alpha D^\beta\psi_\beta+\bar\psi_{\dot{\alpha}}\bar D^{\dot{\alpha}}\bar D^{\dot{\beta}}\bar\psi_{\dot{\beta}})\nonumber\\
&& \ \ \ \ +2\big(1-\frac{1}{\beta}\big)\psi_\alpha D^\alpha\bar D^{\dot{\beta}}\bar\psi_{\dot{\beta}}\big]+2\bar\phi\phi\big] \ ,\\
\label{intver}
&&S_{int}=\frac{1}{2}\int d^8z\big\{(m-8gh\bar\Phi\Phi)V(D^\alpha\psi_\alpha+\bar D^{\dot{\alpha}}\bar\psi_{\dot{\alpha}})+2(2g)\bar\Phi V\phi+2(2g)\Phi\bar\phi V\nonumber\\
&& \ \ \ \ +(2g)^2\bar\Phi\Phi V^2-4h\bar\Phi(D^\alpha\psi_\alpha+\bar D^{\dot{\alpha}}\bar\psi_{\dot{\alpha}})\phi-4h\Phi\bar\phi(D^\alpha\psi_\alpha+\bar D^{\dot{\alpha}}\bar\psi_{\dot{\alpha}})\nonumber\\
&& \ \ \ \ +(2h)^2\bar\Phi\Phi[(D^\alpha\psi_\alpha)D^\beta\psi_\beta+(\bar D^{\dot{\alpha}}\bar\psi_{\dot{\alpha}})\bar D^{\dot{\beta}}\bar\psi_{\dot{\beta}}+2(D^\alpha\psi_\alpha)\bar D^{\dot{\alpha}}\bar\psi_{\dot{\alpha}}]\big\} \ ,
\eea
where the irrelevant terms were omitted, including those involving covariant derivatives of the background (anti-)chiral superfields. Moreover, we used the projection operators $\Pi_{1/2}\equiv-\Box^{-1}D^\alpha\bar D^2D_\alpha$ and $\Pi_{0}\equiv\Box^{-1}\{D^2,\bar D^2\}$.

The one-loop approximation does not depend on how we break the Lagrangian into free and interacting parts \cite{Coleman}. However, by convenience, we will extract the propagators from the terms that are independent of the background superfields and the vertices from the ones in which the quantum superfields interact with the background ones.

In the gauges $\alpha=0$ and $\beta=-1$, we obtain from $S_q$ the propagators
\bea
\label{propagators}
\langle V(1)V(2)\rangle=-\frac{1}{p^2}(\Pi_{1/2})_1\delta_{12} \ , \
\langle\psi_\alpha(1)\bar\psi_{\dot{\alpha}}(2)\rangle=\frac{2p_{\alpha\dot{\alpha}}}{p^4}\delta_{12} \ , \
\langle\phi(1)\bar\phi(2)\rangle=\frac{1}{p^2}\delta_{12} \ .
\eea
Before we start the calculation of the one-loop supergraphs, we first notice from (\ref{intver}) that there is a factor $D^\alpha\bar D^2$ in a vertex at one end of the propagator $\langle\psi_\alpha(1)\bar\psi_{\dot{\alpha}}(2)\rangle$, and there is a factor $\bar D^{\dot{\alpha}}D^2$ in the other vertex at the other end of the same propagator. Here the factors $\bar{D}^2$ and $D^2$ are present in the vertices due to the chirality (antichirality) of the superfield $\psi_{\alpha}$ ($\bar{\psi}_{\dot{\alpha}}$) just as in the usual Wess-Zumino model, because of the properties of the variational derivatives with respect to the chiral superfields (see \cite{SGRS,BuKu,ourcourse}), and the $D^{\alpha}$, $\bar{D}^{\dot{\alpha}}$ arise from the explicit form of the vertices. It is convenient to go from the above used formulation of propagators where the derivatives $D^2$, $\bar{D}^2$ are associated with the vertices, to a formulation where these derivatives are incorporated into the propagators (these two manners to introduce the Feynman supergraphs exist also in the Wess-Zumino model, see f.e. \cite{SGRS}). In other words, we associate the covariant derivatives with the propagator $\langle\psi_\alpha(1)\bar\psi_{\dot{\alpha}}(2)\rangle$ (instead to the vertices) and defining a new scalar field $\psi=D^{\alpha}\psi_{\alpha}$ with the propagator:
\bea
\langle\psi(1)\bar\psi(2)\rangle \equiv D^\alpha_1\bar D^2_1\bar D^{\dot{\alpha}}_2 D^2_2\langle\psi_\alpha(1)\bar\psi_{\dot{\alpha}}(2)\rangle =2(\Pi_{1/2})_1\delta_{12} \ ,
\eea
where we used the fact that $\bar D^{\dot{\alpha}}_2 D^2_2\delta_{12}=-D^2_1\bar D^{\dot{\alpha}}_1\delta_{12}$, and the factors $D^2$, $\bar{D}^2$ emerged due to properties of variational derivatives. We can also apply the same reasoning for the propagator $\langle\phi(1)\bar\phi(2)\rangle$ and for the vertices involving the scalar (anti-)chiral superfields.

In summary, by transferring all covariant derivatives from the vertices (\ref{intver}) to the propagators (\ref{propagators}), we get
\bea
\label{propagator1}
\langle V(1)V(2)\rangle&=&-\frac{1}{p^2}(\Pi_{1/2})_1\delta_{12} \ , \\
\label{propagator2}
\langle\psi(1)\bar\psi(2)\rangle&=&\langle\bar\psi(1)\psi(2)\rangle=2(\Pi_{1/2})_1\delta_{12} \ , \\
\label{propagator3}
\langle\phi(1)\bar\phi(2)\rangle&=&-(\Pi_-)_1\delta_{12} \ , \ \langle\bar\phi(1)\phi(2)\rangle=-(\Pi_+)_1\delta_{12} \ ,
\eea
where $\Pi_-\equiv\Box^{-1}\bar D^2D^2$ and $\Pi_+\equiv\Box^{-1}D^2\bar D^2$ are projection operators. These propagators will connect the following new vertices:
\bea
\label{newintver}
\tilde S_{int}&=&\frac{1}{2}\int d^8z\big\{2MV(\psi+\bar\psi)+2(2g)\bar\Phi V\phi+2(2g)\Phi\bar\phi V+(2g)^2\bar\Phi\Phi V^2\nonumber\\
&&-4h\bar\Phi(\psi+\bar\psi)\phi-4h\Phi\bar\phi(\psi+\bar\psi)+(2h)^2\bar\Phi\Phi[\psi^2+\bar\psi^2+2\psi\bar\psi]\big\} \ ,
\eea
where $M\equiv\frac{1}{2}(m-8gh\bar\Phi\Phi)$. Therefore, now the vertices involve only scalar superfields.

In the next section, we will perform the calculations of the one-loop supergraphs using the propagators (\ref{propagator1}-\ref{propagator3}), written in terms of projection
operators, and the vertices (\ref{newintver}), written only in terms of scalar superfields, instead of the original propagators (\ref{propagators}) and the original vertices (\ref{intver}).

\section{One-loop calculations}

Now, let us start the calculations of the one-loop supergraphs contributing to the KEP. Since $\Pi_{1/2}\Pi_-=\Pi_-\Pi_{1/2}=\Pi_{1/2}\Pi_+=\Pi_+\Pi_{1/2}=0$, it follows from (\ref{propagator1}-\ref{propagator3}) that there can be no mixed contributions containing both gauge and matter propagators at one-loop order. Therefore, the basic supergraphs contributing to the effective action in the theory under consideration are of three types: first, those with internal lines composed of propagators $\langle\psi(1)\bar\psi(2)\rangle$ only; second, those composed of propagators $\langle V(1)V(2)\rangle$ only; third, those involving alternating propagators $\langle\psi(1)\bar\psi(2)\rangle$ and $\langle V(1)V(2)\rangle$.
In our graphical notation, the dashed line is for $<\psi\bar{\psi}>$ propagator, the wavy line is for  $<VV>$ propagator, and the double one is for $\Phi$ or $\bar{\Phi}$ background fields.

\begin{figure}[htbp]
%\begin{center}
%\includegraphics[angle=0,scale=0.55]{spinoroneloop.pdf}
%\end{center}
\centerline{\epsfbox{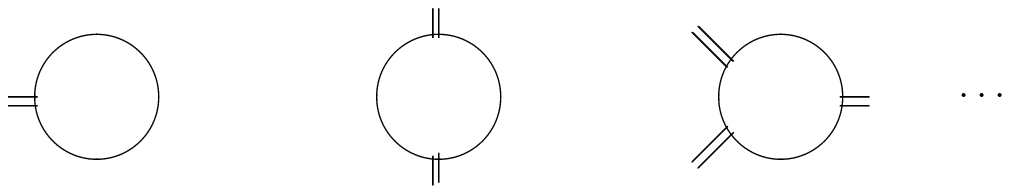}}
\caption{One-loop supergraphs composed by propagators $\langle\psi(1)\bar\psi(2)\rangle$.}
\end{figure}

It is easy to verify that the contribution to the effective action generated by the sum of supergraphs at the Fig. 1, with simple propagators (\ref{propagator2}), and the vertices $2(2h)^2(\Phi\bar{\Phi})\psi\bar{\psi}$ is zero. Indeed, it is equal to
\bea
\Gamma_0=\sum\limits_{n=1}^{\infty}\frac{1}{2n}[4(2h^2)\Phi\bar{\Phi}<\psi\bar{\psi}>]^n,
\eea
where the coefficient 4 is caused by two different contractions. Using the explicit form of the propagators (\ref{propagator2}), we get
\bea
\Gamma_0=\sum\limits_{n=1}^{\infty}\int d^8z_1\frac{1}{2n}[4(2h^2)\Phi\bar{\Phi}\Pi_{1/2}]^n\delta^8(z_1-z_2)|_{z_1=z_2},
\eea
Then, we take into account that $(\Pi_{1/2})^n=\Pi_{1/2}$, and $\Pi_{1/2}\delta^8(z_1-z_2)|_{z_1=z_2}=-2\frac{1}{\Box}\delta^4(x_1-x_2)|_{x_1=x_2}$. Carrying out the Fourier transform, we have
\bea
\Gamma_0=\sum\limits_{n=1}^{\infty}\frac{1}{2n}\int d^8z[4(2h^2)\Phi\bar{\Phi}]^n\int \frac{d^4k}{(2\pi)^4}\frac{1}{k^2},
\eea
but within the dimensional regularization framework implemented through the replacement $d^4k\to \mu^{4-2\omega}d^{2\omega}k$, one has $\int\frac{d^{2\omega}k}{(2\pi)^{2\omega}}\frac{1}{k^2}=0$. Hence, this contribution vanishes.

\begin{figure}[!h]
\begin{center}
\includegraphics[angle=0,scale=0.70]{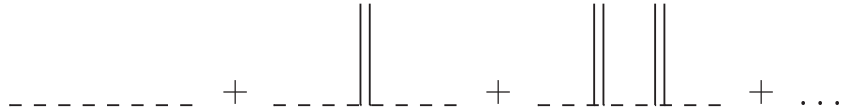}
\end{center}
%\centerline{\epsfbox[angle=0,scale=070]{spinordressedpropagator1.eps}}
\caption{Dressed propagator $\langle \psi(1)\bar\psi(2)\rangle_D$. The vertices are $2(2h)^2(\Phi\bar{\Phi})\psi\bar{\psi}$.}
\end{figure}

Now, let us sum over the vertices $(2h)^2\bar\Phi\Phi\psi^2$ and $(2h)^2\bar\Phi\Phi\bar\psi^2$. The corresponding supergraphs again exhibit structures similar to Fig. 1 with only even number of vertices. However, it is worth to point out that we can insert an arbitrary number of vertices $(2h)^2\bar\Phi\Phi\psi\bar\psi$ into the propagators $\langle\psi(1)\bar\psi(2)\rangle$. Therefore, we should firstly introduce a "dressed" propagator. In this propagator, the summation over all vertices $(2h)^2\bar\Phi\Phi\psi\bar\psi$ is performed (see Fig. 2). As a result, this dressed propagator is equal to
\bea
\label{dressedpropsum}
\langle \psi(1)\bar\psi(2)\rangle_D&=&\langle \psi(1)\bar\psi(2)\rangle+\int d^4\theta_3\langle \psi(1)\bar\psi(3)\rangle[(2h)^2\bar\Phi\Phi]_3\langle \psi(3)\bar\psi(2)\rangle+\int d^4\theta_3d^4\theta_4\nonumber\\
&\times&\langle \psi(1)\bar\psi(3)\rangle[(2h)^2\bar\Phi\Phi]_3\langle \psi(3)\bar\psi(4)\rangle[(2h)^2\bar\Phi\Phi]_4\langle \psi(4)\bar\psi(2)\rangle+\ldots \ .
\eea
By using (\ref{propagator2}), integrating by parts, and summing the resultant series, we arrive at
\bea
\label{dressedpropagator}
\langle \psi(1)\bar\psi(2)\rangle_D=\bigg(\frac{2\Pi_{1/2}}{1-2(2h)^2\bar\Phi\Phi}\bigg)_1\delta_{12} \ .
\eea
Afterwards, we can compute all the contributions by noting that each one-loop supergraph above is formed by $n$ vertices like those ones given by Fig. 3.

\begin{figure}[!h]
\begin{center}
\includegraphics[angle=0,scale=0.625]{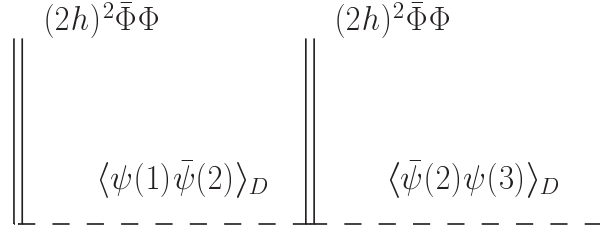}
\end{center}
%\centerline{\epsfbox{spinorlink.eps}}
\caption{A typical vertex in one-loop supergraphs involving $(2h)^2\bar\Phi\Phi\psi^2$ and $(2h)^2\bar\Phi\Phi\bar\psi^2$.}
\end{figure}

Hence, the contribution of this vertex is given by
\bea
Q_{13}&=&\int d^4\theta_2[(2h)^2\bar\Phi\Phi]_1\Big[\Big(\frac{2\Pi_{1/2}}{1-2(2h)^2\bar\Phi\Phi}\Big)_1\delta_{12}\Big][(2h)^2\bar\Phi\Phi]_2\Big[\Big(\frac{2\Pi_{1/2}}{1-2(2h)^2\bar\Phi\Phi}\Big)_2 \delta_{23}\Big]\nonumber\\
&=&\bigg(\frac{2(2h)^2\bar\Phi\Phi}{1-2(2h)^2\bar\Phi\Phi}\Pi_{1/2}\bigg)^2_1\delta_{13} \ .
\eea
It follows from the result above that the contribution of a supergraph formed by $n$ vertices is given by
\bea
I_n&=&\int d^4x\frac{1}{2n}\int d^4\theta_1d^4\theta_3\ldots d^4\theta_{2n-1}\int \frac{d^4p}{(2\pi)^4}Q_{13}Q_{35}\ldots Q_{2n-3,2n-1} Q_{2n-1,1} \nonumber\\
&=&\int d^4x\frac{1}{2n}\int d^4\theta_1d^4\theta_3d^4\theta_5\ldots d^4\theta_{2n-1}\int \frac{d^4p}{(2\pi)^4}\Big[\bigg(\frac{2(2h)^2\bar\Phi\Phi}{1-2(2h)^2\bar\Phi\Phi}\Pi_{1/2}\bigg)^2_1\delta_{13}\Big]\nonumber\\
&\times&\Big[\bigg(\frac{2(2h)^2\bar\Phi\Phi}{1-2(2h)^2\bar\Phi\Phi}\Pi_{1/2}\bigg)^2_3\delta_{35}\Big]\ldots
\Big[\bigg(\frac{2(2h)^2\bar\Phi\Phi}{1-2(2h)^2\bar\Phi\Phi}\Pi_{1/2}\bigg)^2_{2n-1}\delta_{2n-1,1}\Big] \nonumber\\
&=&\int d^8z\frac{1}{2n}\int \frac{d^4p}{(2\pi)^4}\bigg(\frac{2(2h)^2\bar\Phi\Phi}{1-2(2h)^2\bar\Phi\Phi}\bigg)^{2n}\Pi_{1/2}\delta_{\theta\theta^{\prime}}|_{\theta=\theta^{\prime}}\ .
\eea
By using $\Pi_{1/2}\delta_{\theta\theta^{\prime}}|_{\theta=\theta^{\prime}}=2/p^2$, we get the effective action
\bea
\Gamma^{(1)}_1=\sum_{n=1}^{\infty}I_n=-\int d^8z\int \frac{d^4p}{(2\pi)^4}\frac{1}{p^2}\ln\bigg[1-\bigg(\frac{2(2h)^2\bar\Phi\Phi}{1-2(2h)^2\bar\Phi\Phi}\bigg)^{2}\bigg] \ .
\eea
The integral over the momenta vanishes within the dimensional regularization scheme. Therefore,
\bea
\label{part1}
\Gamma^{(1)}_1=0 \ .
\eea
We will not calculate explicitly the one-loop supergraphs involving the gauge superfield propagators $\langle V(1)V(2)\rangle$ connecting the vertices $(2g)^2\bar\Phi\Phi V^2$, because the result is already known and described in \cite{SYM}. Therefore, it is given by
\bea
\label{part2}
\Gamma^{(1)}_2=-\int d^8z\int \frac{d^4p}{(2\pi)^4}\frac{1}{p^2}\ln\bigg[1+\frac{(2g)^2\bar\Phi\Phi}{p^2}\bigg] \ .
\eea

\begin{figure}[!h]
\begin{center}
\includegraphics[angle=0,scale=0.5]{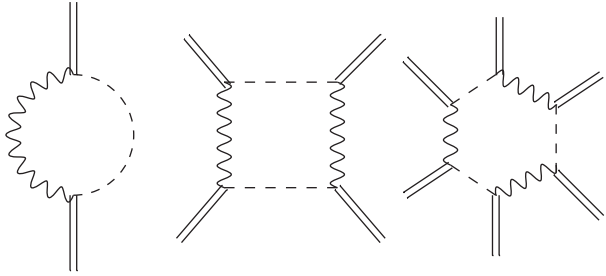}
\end{center}
%\centerline{\epsfbox{mixedoneloop.eps}}
\caption{One-loop supergraphs composed by propagators $\langle\psi(1)\bar\psi(2)\rangle$ and $\langle V(1)V(2)\rangle$.}
\end{figure}
\begin{figure}[!h]
\begin{center}
\includegraphics[angle=0,scale=0.7]{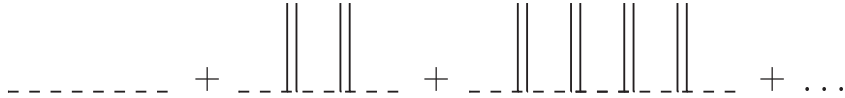}
\end{center}
%\centerline{\epsfbox{spinordressedpropagator2.eps}}
\caption{Dressed propagator $\langle \psi(1)\bar\psi(2)\rangle_{2D}$.}
\end{figure}

Finally, let us move on to the last type of one-loop supergraphs, which involve the propagators $\langle\psi(1)\bar\psi(2)\rangle$ and $\langle V(1)V(2)\rangle$ in the internal lines connecting the vertices $MV\psi$ and $MV\bar\psi$ (see Fig. 4). As before, we can insert an arbitrary number of vertices $(2h)^2\bar\Phi\Phi\psi\bar\psi$ into the propagators $\langle\psi(1)\bar\psi(2)\rangle$. Moreover, we can also insert an arbitrary number of pairs of the vertices $(2h)^2\bar\Phi\Phi\psi^2$ and $(2h)^2\bar\Phi\Phi\bar\psi^2$ into $\langle\psi(1)\bar\psi(2)\rangle$. Since $\langle\psi(1)\bar\psi(2)\rangle$ has already been dressed by $(2h)^2\bar\Phi\Phi\psi\bar\psi$ in (\ref{dressedpropsum}-\ref{dressedpropagator}), it follows that the desired dressed propagator $\langle \psi(1)\bar\psi(2)\rangle_{2D}$ is equal to the summation over all pairs of the vertices $(2h)^2\bar\Phi\Phi\psi^2$ and $(2h)^2\bar\Phi\Phi\bar\psi^2$ into $\langle\psi(1)\bar\psi(2)\rangle_D$ (see Fig. 5). Therefore, we get
\bea
\langle \psi(1)\bar\psi(2)\rangle_{2D}&=&\langle \psi(1)\bar\psi(2)\rangle_D+\int d^4\theta_3d^4\theta_4\langle \psi(1)\bar\psi(3)\rangle_D[(2h)^2\bar\Phi\Phi]_3\langle \bar\psi(3)\psi(4)\rangle_D\nonumber\\
&\times&[(2h)^2\bar\Phi\Phi]_4\langle \psi(4)\bar\psi(2)\rangle_D+\int d^4\theta_3d^4\theta_4d^4\theta_5d^4\theta_6\langle \psi(1)\bar\psi(3)\rangle_D[(2h)^2\bar\Phi\Phi]_3\nonumber\\
&\times&\langle \bar\psi(3)\psi(4)\rangle_D[(2h)^2\bar\Phi\Phi]_4\langle \psi(4)\bar\psi(5)\rangle_D[(2h)^2\bar\Phi\Phi]_5\langle \bar\psi(5)\psi(6)\rangle_D\nonumber\\
&\times&[(2h)^2\bar\Phi\Phi]_6\langle \psi(6)\bar\psi(2)\rangle_D+\ldots \ .
\eea
After some algebraic work, we find
\bea
\langle \psi(1)\bar\psi(2)\rangle_{2D}=(2f(\bar\Phi\Phi)\Pi_{1/2})_1\delta_{12} \ , \ \textrm{where} \ f(\bar\Phi\Phi)\equiv\frac{1}{1-4(2h)^2\bar\Phi\Phi} \ .
\eea
Additionally, we can also insert an arbitrary number of vertices $(2g)^2\bar\Phi\Phi V^2$ into the propagators $\langle V(1)V(2)\rangle$. In this case, the dressed propagator $\langle V(1)V(2)\rangle_D$ is already known in the literature and it is given by \cite{Our}
\bea
\label{Vdressedpropagator}
\langle V(1)V(2)\rangle_D=\bigg(\frac{-\Pi_{1/2}}{p^2+(2g)^2\bar\Phi\Phi}\bigg)_1\delta_{12} \ .
\eea
As before, we can compute all the contributions by noting that each supergraph above (Fig. 4) is formed by $n$ fragments, like those depicted in Fig. 6. This fragment yields the contribution
\bea
R_{13}&=&\int d^4\theta_2(M)_1\Big[\Big(\frac{-\Pi_{1/2}}{p^2+(2g)^2\bar\Phi\Phi}\Big)_1\delta_{12}\Big](M)_2\Big[(2f\Pi_{1/2})_2\delta_{23}\Big]\nonumber\\
&=&\bigg(\frac{-2fM^2\Pi_{1/2}}{p^2+(2g)^2\bar\Phi\Phi}\bigg)_1\delta_{13} \ .
\eea

\begin{figure}[!h]
\begin{center}
\includegraphics[angle=0,scale=0.625]{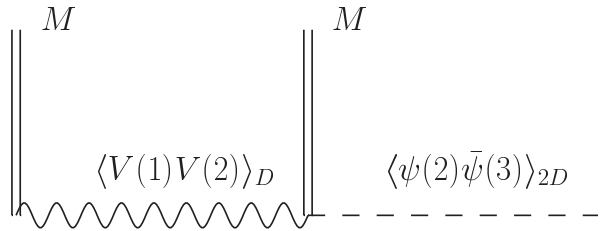}
\end{center}
%\centerline{\epsfbox{mixedlink.eps}}
\caption{A typical vertex in one-loop supergraphs involving $MV\psi$ and $MV\bar\psi$.}
\end{figure}

It follows from the result above that the contribution of a supergraph formed by $n$ subgraphs is given by
\bea
J_n&=&\int d^4x\frac{1}{2n}\int d^4\theta_1d^4\theta_3\ldots d^4\theta_{2n-1}\int \frac{d^4p}{(2\pi)^4}R_{13}R_{35}\ldots R_{2n-3,2n-1} R_{2n-1,1} \nonumber\\
&=&\int d^4x\frac{1}{2n}\int d^4\theta_1d^4\theta_3d^4\theta_5\ldots d^4\theta_{2n-1}\int \frac{d^4p}{(2\pi)^4}\Big[\bigg(\frac{-2fM^2\Pi_{1/2}}{p^2+(2g)^2\bar\Phi\Phi}\bigg)_1\delta_{13}\Big]\nonumber\\
&\times&\Big[\bigg(\frac{-2fM^2\Pi_{1/2}}{p^2+(2g)^2\bar\Phi\Phi}\bigg)_3\delta_{35}\Big]\ldots
\Big[\bigg(\frac{-2fM^2\Pi_{1/2}}{p^2+(2g)^2\bar\Phi\Phi}\bigg)_{2n-1}\delta_{2n-1,1}\Big] \nonumber\\
&=&\int d^8z\frac{1}{2n}\int \frac{d^4p}{(2\pi)^4}\bigg(\frac{-2fM^2}{p^2+(2g)^2\bar\Phi\Phi}\bigg)^n\Pi_{1/2}\delta_{\theta\theta^{\prime}}|_{\theta=\theta^{\prime}}\ .
\eea
Again, by using $\Pi_{1/2}\delta_{\theta\theta^{\prime}}|_{\theta=\theta^{\prime}}=2/p^2$, we get the effective action
\bea
\label{part3}
\Gamma^{(1)}_3=\sum_{n=0}^{\infty}J_n=-\int d^8z\int \frac{d^4p}{(2\pi)^4}\frac{1}{p^2}\ln\bigg[1+\frac{2fM^2}{p^2+(2g)^2\bar\Phi\Phi}\bigg] \ .
\eea
By summing (\ref{part1}), (\ref{part2}), and (\ref{part3}) we obtain the total one-loop effective action
\bea
\label{totalEA}
\Gamma^{(1)}[\bar\Phi,\Phi]=-\int d^8z\int \frac{d^4p}{(2\pi)^4}\frac{1}{p^2}\ln\big[p^2+(2g)^2\bar\Phi\Phi+2fM^2\big] \ .
\eea
Substituting the explicit form for $M$ and $f$, we arrive to the following result for the KEP:
\bea
\label{totalEP}
K^{(1)}(\bar\Phi,\Phi)=-\int \frac{d^4p}{(2\pi)^4}\frac{1}{p^2}\ln\big[p^2+\frac{1}{2-8(2h)^2\bar\Phi\Phi}(m-8gh\bar\Phi\Phi)^2+(2g)^2\bar\Phi\Phi\big] \ .
\eea

The integral above is well-known and can be computed by using the dimensional regularization. Finally, in the limit $\omega\rightarrow2$ we find
\bea
\label{mainresult}
K^{(1)}(\bar\Phi,\Phi)=K_{div}^{(1)}(\bar\Phi,\Phi)+K_{fin}^{(1)}(\bar\Phi,\Phi) \ ,
\eea
where
\bea
\label{divresult}
K_{div}^{(1)}(\bar\Phi,\Phi)&=&\frac{1}{16\pi^2(2-\omega)}\Big[\frac{1}{2-8(2h)^2\bar\Phi\Phi}(m-8gh\bar\Phi\Phi)^2+(2g)^2\bar\Phi\Phi\Big] \ ,\\
\label{finresult}
K_{fin}^{(1)}(\bar\Phi,\Phi)&=&-\frac{1}{16\pi^2}\Big[\frac{1}{2-8(2h)^2\bar\Phi\Phi}(m-8gh\bar\Phi\Phi)^2+(2g)^2\bar\Phi\Phi\Big]\nonumber\\
&\times&\ln\frac{1}{\mu^2}\Big[\frac{1}{2-8(2h)^2\bar\Phi\Phi}(m-8gh\bar\Phi\Phi)^2+(2g)^2\bar\Phi\Phi\Big] \ ,
\eea
and $\mu$ is an arbitrary scale required on dimensional grounds. 

Notice that the one-loop KEP (\ref{mainresult}-\ref{finresult}) is divergent. Moreover, we notice that the divergent part (\ref{divresult}) is given by an infinite power series in $\bar\Phi\Phi$. Therefore, the theory under consideration is non-renormalizable and it must be interpreted as an effective field theory below some energy scale chosen on the basis of phenomenological considerations \cite{Bur}.

In particular, let us take $h=0$ in (\ref{mainresult}). This choice corresponds to a minimal coupling between the gauge scalar superfield and the matter chiral superfields [see (\ref{matteract})]. Therefore,
\bea
\label{particulardiv}
K_{div}^{(1)}(\bar\Phi,\Phi)&=&\frac{(2g)^2\bar\Phi\Phi}{16\pi^2(2-\omega)} \ ,\\
\label{particularfin}
K_{fin}^{(1)}(\bar\Phi,\Phi)&=&-\frac{1}{32\pi^2}\big[m^2+2(2g)^2\bar\Phi\Phi\big]\ln\frac{1}{2\mu^2}\big[m^2+2(2g)^2\bar\Phi\Phi\big] \ .
\eea
In this case, we notice that the divergent term (\ref{particulardiv}) is proportional to $\bar\Phi\Phi$. Therefore, in order to remove divergences, we can insert a similar one-loop counterterm as the one used in the SQED. Moreover, if we take the massless case in (\ref{particulardiv}-\ref{particularfin}), we recover the one-loop KEP for the usual SQED \cite{SYM}.

\section{Summary}

We formulated a new theory involving coupling of three superfields of different natures: a chiral spinor gauge superfield originally introduced in \cite{Siegel} together with the usual real scalar gauge superfield and the chiral scalar matter superfield. For this theory, we developed a superfield procedure for calculating the one-loop effective potential which we successfully found. The procedure does not essentially differ from the usual supergauge theories \cite{SYM} with the rather similar structure of the one-loop contribution. The fact that the new theory is non-renormalizable is not unexpected since many non-polynomial supersymmetric theories are non-renormalizable \cite{GC,Brignole}. We expect that this theory, besides of the classical studies  in the cosmic string context, can be used as an ingredient of possible phenomenologically interesting supersymmetric gauge theories involving several gauge (super)fields with some of them being massive.

\vspace*{3mm}

{\bf Acknowledgments.} This work was partially supported by Conselho
Nacional de Desenvolvimento Cient\'{\i}fico e Tecnol\'{o}gico (CNPq). The work by A. Yu. P. has been partially supported by the
CNPq project No. 303438/2012-6.

\end{document}